\documentclass[a4paper,12pt]{spieman}  
\usepackage{amsmath,amsfonts,amssymb}
\usepackage{graphicx}
\graphicspath{{figs/}{}}
\hypersetup{pdfstartview={FitH},pdfpagemode={UseNone},
            colorlinks,linkcolor=blue, citecolor=blue, urlcolor=blue,
            bookmarksopen=true, pdfnewwindow=true}
\usepackage[all]{hypcap}

\usepackage{setspace}
\usepackage{tocloft}
\usepackage[table]{xcolor}

\usepackage{soul}
\usepackage{array}
\newcolumntype{x}[1]{>{\raggedright}p{#1}}

\usepackage{tabulary}

\usepackage{pdfcomment}
\usepackage[colorinlistoftodos]{todonotes}
\usepackage{siunitx}
\usepackage{textcomp}
\usepackage{bm}
\usepackage{multirow}
\usepackage{lineno}

\title{Metasurface-based Toroidal Lenslet Array Design for Addressing Laser Guide Star Elongation}

\author[a,*]{Josephine Munro} 
\author[a,*]{Sarah E. Dean} 
\author[a]{Neuton Li}
\author[b]{Israel J. Vaughn} 
\author[b]{Andrew W. Kruse} 
\author[b]{Tony~Travouillon} 
\author[a]{Dragomir N. Neshev} 
\author[b]{Robert Sharp} 
\author[a]{Andrey A. Sukhorukov} 
\affil[a]{ARC Centre of Excellence for Transformative Meta-Optical Systems (TMOS), Department of Electronic Materials Engineering, Research School of Physics, Australian National University, Canberra, ACT 2600, Australia}
\affil[b]{Research School of Astronomy and Astrophysics, Australian National University, Weston Creek, ACT 2611, Australia}

\cftpagenumbersoff{figure}
\cftpagenumbersoff{table} 

\hypersetup{pdfstartview={FitH},pdfpagemode={UseNone}, breaklinks=true,
            colorlinks,linkcolor=blue, citecolor=blue, urlcolor=blue,
             bookmarksopen=true, pdfnewwindow=true}
\usepackage[all]{hypcap}
\newcommand*{\email}[1]{\href{mailto:#1}{\nolinkurl{#1}} }

\begin{document} 
\sloppy
\maketitle

\begin{abstract}
The Giant Magellan Telescope will use laser tomography adaptive optics to correct for atmospheric turbulence using artificial guide stars created in the sodium layer of the atmosphere (altitude $\approx$~\qty{95}{km}). The sodium layer has appreciable thickness ($\approx$~\qty{11}{km}) and this results in the laser guide star being an elongated cylinder shape. Wavefront sensing with a Shack-Hartmann is challenging, as subapertures located further away from the laser launch position image an increasingly elongated perspective of the laser guide star. Large detectors can be used to adequately pack and sample the images on the detector, however, this increases readout noise and limits the design space available for the wavefront sensor. To tackle this challenge, we propose an original solution based on nano-engineered meta-optics tailored to produce a spatially varying anamorphic image scale compression. We present
meta-lenslet array designs that can deliver $\approx$100\% of the full anamorphic image size reduction required for focal lengths down to 8 mm, and greater than 50\% image size reduction for focal lengths down to 2 mm. This will allow greatly improved sampling of the available information across the whole wavefront sensor, while still being a viable design within the limits of current-generation fabrication facilities.
\end{abstract}

\keywords{meta-lenslet array, Shack-Hartmann wavefront sensor, laser guide star, anamorphic image scale compression}

{\noindent \footnotesize\textbf{*}Equal contribution from lead authors Josephine Munro,  \email{josephine.munro@anu.edu.au}\newline and Sarah Dean, \email{sarah.dean@anu.edu.au} }

\section{\label{sec:Introduction}Introduction}

With a primary mirror diameter of 24.5~m, The Giant Magellan Telescope (GMT~\cite{Franson2020}) is one of three next-generation ground-based Extremely Large Telescopes (ELTs) scheduled to begin operations in the next decade. As with all large ground-based optical telescopes, Adaptive Optics (AO) will be used to correct atmosphere-induced aberrations to ensure the ELTs deliver images at or near the diffraction limit. AO requires a sufficiently bright star in close angular proximity to the science target of interest to provide a reference for accurate wavefront correction~\cite{Lombini2021}. Although natural guide stars offer a high quality of correction, there is not always a bright star sufficiently close to the target of interest, which restricts sky coverage for what can and cannot be observed with AO systems~\cite{Ellerbroek2008}. This can be overcome with the creation of an artificial laser guide star (LGS) close to the science target. A laser with a wavelength resonant to the spectral doublet, known as the Sodium D-lines, is launched from near the telescope to excite sodium atoms at the altitude of the mesosphere sodium layer ($\approx$~\qty{90}{km})~\cite{Pfrommer2014}. The excited sodium atoms resonance fluoresce; releasing a \qty{589}{nm} photon as they relax from the excited $3p$ into the lower $3s$ energy state~\cite{Lihang2016}. This luminescent emission is the artificial LGS. Owing to their ability to drastically increase sky coverage, LGS facilities will be available at all three extremely large telescopes.

The GMT will have three AO modes~\cite{Bouchez2014}; Natural Guide Star Adaptive Optics, Laser Tomography Adaptive Optics (LTAO), and Ground Layer Adaptive Optics. Of particular interest to this study is the LTAO mode which uses an asterism of six LGS and one off-axis natural guide star to achieve $<$~\qty{290}{nm} root-mean-square (RMS) wavefront error with $>$~\qty{50}{\%} sky coverage at the galactic pole for near-infrared diffraction-limited imaging~\cite{Bouchez2014}. The six laser launch locations are peripherally positioned around the primary mirror. The wavefront sensor for the LTAO system will be a Shack-Hartmann design~\cite{Conan2012}, which is the standard choice of wavefront sensors for ELTs due to their simplicity and well-characterized operation~\cite{Lombini2021}. Despite this, Shack-Hartmann wavefront sensors (SHWFS) face substantial challenges when trying to sample the wavefront optimally due to laser guide star elongation; a phenomenon that will especially affect ELTs.  

\subsection{Laser Guide Star Elongation}

\begin{figure}
    \centering
    \includegraphics[width=0.9\textwidth]{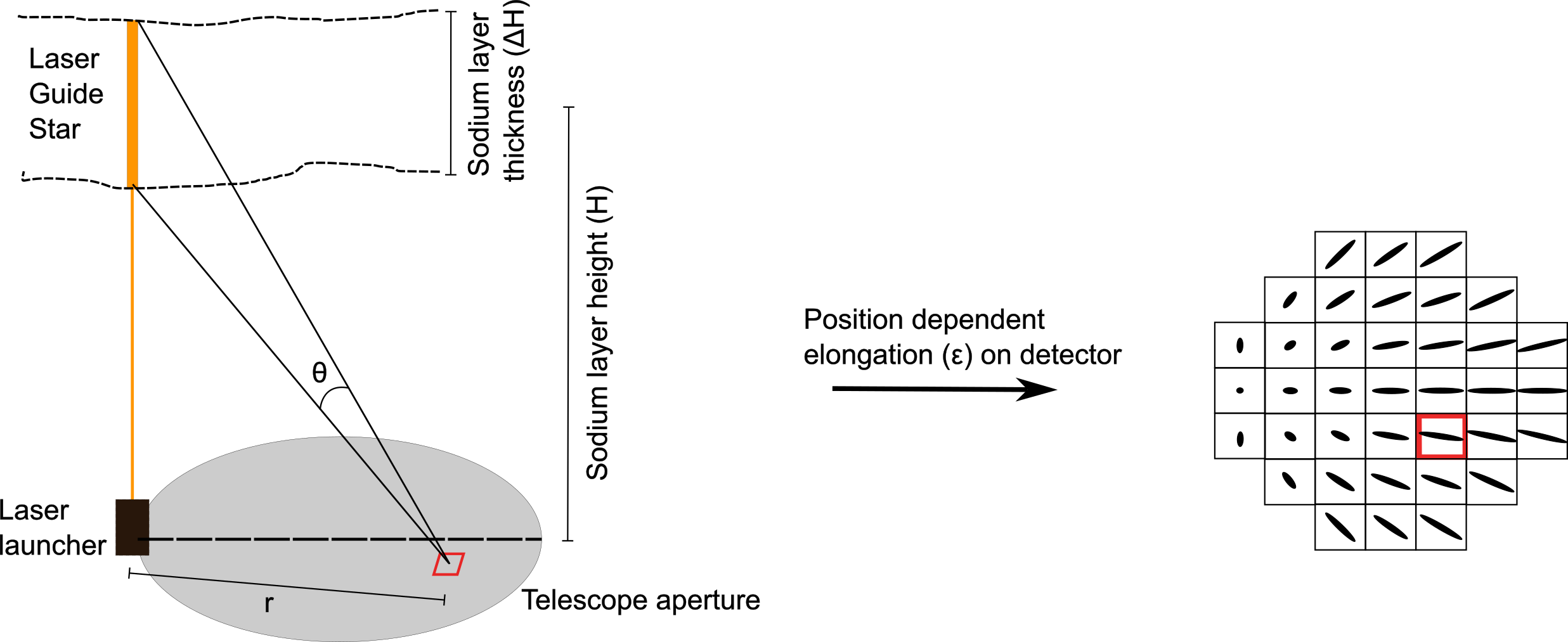}
    \caption{Schematic of laser guide star elongation due to the finite thickness of the sodium layer in the atmosphere. The perspective elongation in each subaperture increases as its distance from the laser launch position increases. Similarly, the angle of the elongated axis is azimuthally dependent on each subaperture's position with respect to the laser launch position.}
    \label{fig:LGS}
\end{figure}

The atmospheric sodium layer is not infinitely thin; the vertical extension (thickness) is approximately \qtyrange{10}{20}{\kilo\metre}~\cite{Pfrommer2014,Neichel2013}. The volumetric shape of the LGS is a column of illumination along the laser beam axis, as opposed to an ideal point source. This has the effect of the artificial star appearing elongated when observed from a laterally offset position with respect to the laser launcher~\cite{Lombini2021}, as geometrically depicted in Figure~\ref{fig:LGS}. The angular image elongation, $\varepsilon$, can be approximated for each subaperture position as
\begin{equation}
    \varepsilon(r) \approx r \frac{\Delta H}{H^2} \cos(\theta) ,
    \label{eq:elongation}
\end{equation}
where $r$ is the distance of the subaperture (imposed on the primary mirror) from the laser launch position, $\Delta H$ is the thickness of the sodium layer, $H$ is the height of the sodium layer, and $\theta$ is the on-sky angle subtended by the LGS in the subaperture~\cite{Lombini2021}. 

Subapertures with larger lateral displacement to the laser launcher have a larger image elongation. On the GMT, for example, the maximum elongation will be $\varepsilon \approx$~\qty{9.7}{arcsec} for a subaperture on the opposite edge of the pupil to the laser launch position. The image size in the directional orthogonal to the elongation is considered to be \qty{1}{arcsec}~\cite{Conan2012}. The maximum elongation ratio of the LGS image for ELTs is between 1:9 to 1:13 depending on the telescope's diameter and the atmospheric statistics of each telescope site.

Perspective elongation is challenging because it makes packing the images onto the detector difficult. Along with the normal design-space constraints of sensitivity and dynamic range, further consideration needs to be taken to avoid the overlap of neighbouring images and avoid excessive truncation of the most elongated images. The standard solution to address image elongation has been to compromise the detector plate scale. Normally, the plate scale can be chosen such that the trade-off between the sampling and the field of view (FOV) still satisfies system requirements. But to combat elongation, the FOV should be maximized to avoid excessive truncation, simultaneously the sampling of non-elongated images must be adequate for the centroiding algorithm used, given the noise characteristics of the system. This is achieved by using a large number of pixels to sample a large FOV in each subaperture, resulting in large-area detectors~\cite{Fusco2022}. However, this approach generally increases the readout noise, the readout time, the computational time and the power required. Furthermore, a large amount of the detector is unused; sampling blank sky around the less elongated spots and in the transverse direction to elongated ones. Fixing a minimum subaperture FOV to limit the bias that is introduced by truncation of the elongated images reduces the available design space of the wavefront sensor. For example, the compromise of the plate scale for GMT's LTAO leads to 0.71~arcsec per pixel in a system expecting 1~arcsec non-elongated LGS image size~\cite{Conan2012}. 

This problem has attracted multiple proposals that aim to minimize the effect of LGS elongation, with solutions that are computational, detector, and optically based. Computational-based solutions include discarding or weighing the elongation-axis wavefront information from each LGS in an asterism and calculating the missing information from a combination of the other signals~\cite{Fusco2019}. Detector-based approaches include variable gains for different regions of interest~\cite{Downing2012}, and CCDs with custom pixel morphology (where pixels in each subaperture are radially aligned to the elongation axis)~\cite{Beletic2005}. However, this approach still requires a large number of pixels to limit truncation to an acceptable level. 

Optical domain solutions have been previously suggested, including combinations of multiple arrays of custom optics, free-form solutions~\cite{Jahn2016}, and tilting prism arrays~\cite{Gendron2016}, but have not been investigated beyond theoretical and modelling stages of development. A viable optical solution has to be able to address the elongation in each subaperture while meeting requirements for fabrication, integration and alignment. There has been recent renewed interest in finding an optical domain solution to the elongation problem with aspirations of increasing the sensitivity, accuracy and efficiency of AO systems for ELTs. In the context of the Thirty Meter Telescope, Lombini et al.,~\cite{Lombini2021} detailed the design and simulation results of a bi-prism lenslet array, originally described by Schreiber et al.~\cite{Schreiber2008}. Each subaperture bi-prism is aligned along the axis of elongation, effectively acting as a pyramid prism with only two faces. The elongated signal is divided into two images of the pupil, where the difference in intensity between the two images is proportional to the local tip/tilt in the orthogonal axis to the elongation. This method is insensitive to the aberrations occurring in the elongation axis. A recombination algorithm using other LGS signals could be used to reconstruct the elongation axis aberrations. The recombination process is not trivial and is another source of wavefront error. Notwithstanding, the bi-prism array is a promising optical solution as simulations have shown it allows for reducing the size of the detector; from 14-20 pixels$^2$ per subaperture to 2-6 pixels$^2$ per subaperture. 

Combating the negative effects of elongation is an important, though challenging, problem. It would be ideal to set a variable plate scale across the detector to best sample the available information in each subaperture. Anamorphic compression is widely done in the optics community with a pair of cross cylindrical lenses. However, parfocal operation across an array requiring variable anamorphic compression with only a pair of crossed cylindrical lenslets is not possible since either the back focal distance or the air gap between the cylindrical lenslets would need to vary across the array. Taking the principle of a toroidal lens, which has different focal lengths in orthogonal axes, two separate layers of lenslets arrays have enough degrees of freedom to achieve parfocal operation and satisfy a large variation in anamorphic compression. To the authors' knowledge, there has been no significant progress in the fabrication of a variably anamorphic lenslet array comprised of crossed cylinders, prism pairs, or toroidal lenses using conventional optics. There have been preliminary fabrication efforts towards 3D diamond machining of free-form micro-optics designs which includes working towards spatially varying lenslet sizes and dithered lenslets. However, it is noted that there are significant limitations to the capability of the fabrication technique due to the finite tool size setting a limit for the size of discontinuities in the workpiece~\cite{Schmoll2006}.   

\begin{figure}
    \centering
    \includegraphics[width=\textwidth]{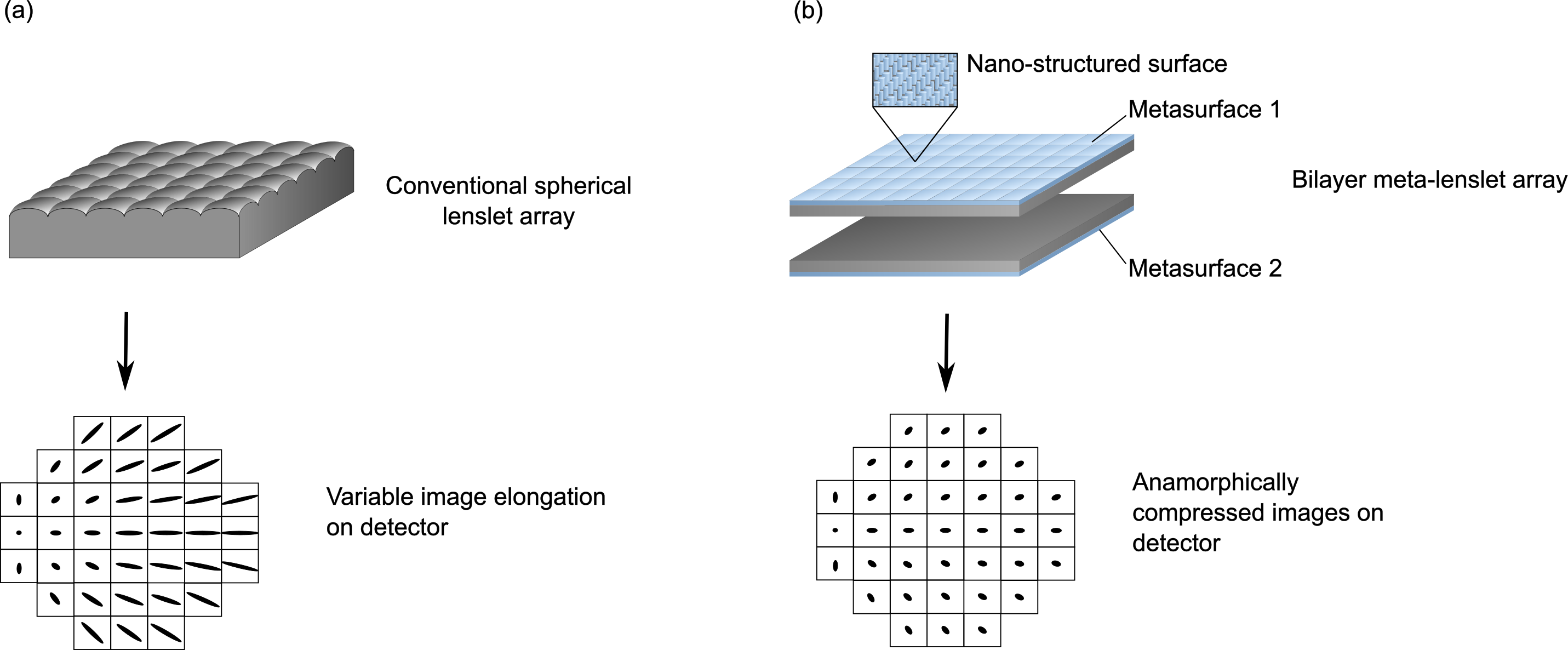}
    \caption{(a) Laser guide star image elongation on a WFS detector with a conventional spherical lenslet array. The images in subapertures that are laterally displaced further away from the laser launch position have a larger elongation ratio. (b) Anamorphically compressed images on a WFS detector with our proposed meta-lenslet array. Two layers of metasurfaces (bilayer) are fabricated on a fused silica substrate and separated by an air gap. With the shrinking of the image size across the whole aperture, the number of pixels required in each subaperture can be reduced, reducing the size of the detector.}
    \label{fig:LensletArray2}
\end{figure}

Conversely, a variable plate scale is readily achievable with the meta-lenslet approach presented here. We propose a meta-lenslet array that is designed to individually address the elongation ratio and azimuthal angle in each subaperture, as depicted in Figure~\ref{fig:LensletArray2}(b). The field of metaoptics has grown rapidly, with instrumentation applications evolving alongside further fundamental nanophotonics research~\cite{Neshev:2023-26:NPHOT}. Nanostructured flat metaoptics have a wide range of applications that are particularly suited to optical systems with constraints on size, weight, and power requirements. In the literature, there are examples of toroidal meta-lenses in a zoom telescope system that have been shown to enable variable magnification when switching between orthogonal axes~\cite{Bregenzer2020,Yang2022}. However, metaoptics have never been explored for LGS elongation correction before. We show how this meta-lenslet array solution is viable, with the parfocal operation across a broad range of anamorphic compression ratios achieved with two layers (a bilayer) of metasurfaces separated by an air gap.  

\subsection{\label{subsec:MSlit}Meta-lenslet Approach} 

Conventional optics rely on propagation through bulk media to manipulate the optical properties of the incoming light, whereas metasurfaces induce an abrupt change in optical properties at an ultra-thin surface~\cite{Chen2016}. Metasurfaces are artificially designed and fabricated arrays of sub-wavelength sized structures that are purpose-designed to manipulate the incoming wavefront to achieve a desired output wavefront, for example with a particular amplitude, phase, or polarisation~\cite{Khorasaninejad2017}. In infrared and visible spectral range of operation, the size of the sub-wavelength-sized structures is on the order of tens to hundreds of nanometres, leading to the term `nanostructures'. Geometric parameters such as the size, shape and orientation of each nanostructure determine the local effect on the wavefront. By varying a geometric parameter across the metasurface, the output wavefront can be controlled. The manipulation of the incoming light can replicate the functions of conventional optics, but can also go beyond what conventional bulk optics can achieve, especially for multi-mode and multi-functional capabilities. The use of dielectric instead of metallic materials was an important advancement in the field; as materials such as silicon (Si) and numerous silicon compounds, gallium nitride ($\mathrm{GaN}$), and titanium dioxide ($\mathrm{TiO_2}$) have significantly less absorption loss than metallic materials, making them suitable for use with metasurfaces in the infrared and visible regions. Additionally, the high refractive index of dielectric materials allows for better light confinement and reduced near-field coupling. The behaviour of the dielectric structures can be described by the interference of supported Mie-like resonances within the structure~\cite{Koshelev2021}. Metasurfaces can be fabricated using existing micro and nanofabrication technologies commonly used in the semiconductor manufacturing industry. 

Both cylindrical and spherical focusing meta-lenses are commonly designed by determining the phase profile required to manipulate an incident wavefront, and then arranging nanoresonator elements from a library with predetermined phase transmissions to match the target phase profile. To focus a planar wavefront at normal incidence, a metasurface should impose a hyperbolic phase delay on the incoming wavefront to achieve aberration-free focusing~\cite{Pan2022}. Similarly, a cylindrical metalens also imposes a hyperbolic phase delay but is only spatial dependent on one axis. This concept is easily extended to anamorphic behaviour by introducing coefficients ${a_1}$ and ${a_2}$ for $x$ and $y$, respectively, as follows:
    \begin{equation}
        \psi (x,y) = \frac{2 \pi}{\lambda}\left( f - \sqrt{a_1 x^2 + a_2 y^2 + f^2} \right) ,
        \label{eq:ana}
    \end{equation}
where $f$ is the focal length of the metalens, and ${a_1} = {a_2}$ is the special case of spherical focusing and ${a_1}=0$ or ${a_2} = 0$ is the special case of cylindrical focusing. Applications for spherical and cylindrical meta-lenses include imaging and microscopy~\cite{Bayati2020,Khorasaninejad2016,Arbabi2016,Wen2016}. Additional numerical optimization is often used to correct aberrations or improve performance. The metasurface materials, thickness, and resonator shapes are all tailored to the operating wavelengths and behaviours desired from the metasurface~\cite{Pan2022}. A more complex design was demonstrated by Li et al.~\cite{Li2022}, where a metasurface for polarization-dependent beam steering utilized anamorphic metalenses. The anamorphic lens components were also aspheric to properly function as part of the beam-steering doublet, and were implemented by extracting the phase profiles from a designed geometric doublet and discretizing it appropriately for individual nanoresonators. The maximum ratio of focal lengths between the $x$ and $y$ directions for the designed anamorphic lenses demonstrated by Li et al.~\cite{Li2022} was approximately $1:3.5$.

Metasurface-based replication of lenslet arrays has focused on instrument advances towards miniaturization~\cite{Liu2019}, or additional functionality such as polarisation sensitivity for beam profiling~\cite{Yang2018}. For the application to LGS elongation, metasurfaces with polarisation independence will be desirable. Polarisation independence is typically achieved by using nanoresonators with three or higher degrees of symmetry in the $x,y$ plane, such as cylindrical nanopillars, which avoids the shape birefringence.
The lattice arrangement should also have a high degree of symmetry to avoid lattice-induced birefringence. Polarisation-insensitive metasurfaces are well-represented in previous literature, with several examples existing specifically for metalenses~\cite{Arbabi2016, Vo2014, Khorasaninejad2016b}. A polarisation independent 60x60 meta-lenslet array was fabricated for long-wave infrared (\qty{10.6}{\um}),  with a pitch size of \qty{100}{\um} and focal length of \qty{100}{\um}, with good agreement between simulations of focal spot size and laboratory measurements~\cite{Liu2019}. A significant drop from the simulated efficiency of 43\% to the measured efficiency of 34\% was noted. The authors postulate that reflection from the substrate backside and the absorbance of the silicon substrate at \qty{10.6}{\um} are major contributing factors to the efficiency loss, and so the efficiency could be improved by employing an anti-reflective coating, and choosing a substrate material with a higher long-wave infrared transparency. 

The above examples of metasurface-based lenslet arrays have been designed for infrared wavelengths, however, our LGS application will be specifically for a wavelength of \qty{589}{nm} in the visible region. Metasurfaces operating at visible wavelengths use materials such as $\mathrm{TiO_2}$~\cite{Khorasaninejad2016} and GaN~\cite{Wang2018} for increased efficiency compared to silicon. However, a material analysis study showed crystalline silicon (c-Si), for a \qty{500}{nm} thick metasurface structure, performs similarly to $\mathrm{TiO_2}$ for a \ang{70} deflection meta-grating at \qty{550}{nm}~\cite{Yang2017}.

An analytical study quantified the cross-talk prevalence for small pitch-sized $\mathrm{TiO_2}$ on fused silica substrate meta-lenslet arrays at \qty{532}{nm}. Their findings showed pitch sizes larger than \qty{8}{\um} and dead-space between meta-lenslets larger than \qty{0.5}{\um} resulted in negligible cross-talk between meta-lenslets~\cite{Ozdemir2018}. This robust numerical study provides a sound technical reference to designing a meta-lenslet array for visible wavelengths, with the caveat that experimental results are required to confirm the characteristic optical behaviour and to characterise the effect of noise and other system errors that are otherwise ignored for simplicity in simulations. To the authors' knowledge, this study presents the first design of a metasurface-based array to directly and uniquely minimize the effect of LGS elongation for SHWFS.


\section{Meta-lenslet Optical System Modelling}
\label{sec:OpticalDesigns}

The lenslet parameter values of the LTAO WFS for the GMT~\cite{Conan2012} were used as the design reference for the meta-lenslet optical design, and are given in Table~\ref{tab:ReqPar}. 

\begin{table}
    \caption{Meta-lenslet design values}
    \centering
    \begin{tabular}{l l} \hline \hline
    \rule[-1ex]{0pt}{2ex} \textbf{Parameter} & \textbf{Value} \\ \hline 
    \rule[-1ex]{0pt}{2ex} Effective focal length & \qty{2}{mm} \\ \hline
    \rule[-1ex]{0pt}{2ex} Pitch size (meta-lenslet diameter) & \qty{400}{\um} \\ \hline 
    \rule[-1ex]{0pt}{2ex} Wavelength & \qty{589}{nm} \\ \hline
    \rule[-1ex]{0pt}{2ex} Range of elongation ratios & 1:1 - 1:10  \\ \hline \hline
    \end{tabular}
    \label{tab:ReqPar}
\end{table}  

Maintaining the effective focal length for the non-elongated axis preserves the WFS characteristics in the GMT's LTAO design. The effective focal length in the non-elongated axis (hereafter denoted $f_{\alpha}$) is held constant across all meta-lenslets in the array. To achieve anamorphic compression in paraxial conditions, the effective focal length in the elongated axis (hereafter denoted as $f_{\beta}$) is smaller than $f_{\alpha}$ by a factor equal to the anamorphic ratio. $f_{\beta}$ therefore varies across the meta-lenslet array. If we consider paraxial conditions for the case of 1:10 compression, and $f_{\alpha}$ = \qty{2}{mm}, then $f_{\beta}$ = \qty{0.2}{mm}, which gives a numerical aperture of 1 in the elongated axis. Large numerical apertures are difficult to achieve as they require sophisticated full-scale optimization designs~\cite{phan2019high}. We limited our $\mathrm{2\pi}$ phase shift gradient to \qty{3}{\um} linear distance on the meta-lenslet to not exceed the typical fabrication capabilities. Consequently, the shortest focal length achievable in our simulations will be $\approx$ \qty{0.8}{mm} (numerical aperture $\approx$~0.25). This is not to imply the problem is unsolvable by metasurfaces; much higher numerical apertures have been successfully demonstrated but are more complex to fabricate, and have trade-offs with increasing aberrations~\cite{Liang2019}. 

The paraxial numerical aperture in the elongated axis is tabulated in Table~\ref{tab:NAf} for our design target of $f_{\alpha}$ = \qty{2}{mm}, as well as longer focal lengths up to $f_{\alpha}$ = \qty{10}{mm}. It is clear from Table~\ref{tab:NAf} that most of the anamorphic ratios for $f_{\alpha}$ = \qty{2}{mm} will require a numerical aperture in the elongation axis above our achievable limit. We, therefore, expect to have less than \qty{100}{\percent} reduction in the image size in subapertures with numerical apertures in the elongation axis larger than 0.25. It should be noted that even a reduction of \qty{50}{\percent} in image size would still significantly improve the LGS WFS. GMT's LTAO design value of $f_\alpha$ = \qty{2}{mm} is unusually low to maximize FOV to avoid excessive truncation. This is why we also modelled effective focal lengths up to \qty{10}{mm}, investigating a representative range of common lenslet array focal lengths. With ten individual meta-lenslets designed with anamorphic ratios from 1:1–1:10 for each focal length between 2-\qty{10}{mm}; a total of 90 meta-lenslets were designed.

\begin{table}
    \centering
    \caption{Paraxial solution numerical aperture in the elongated axis for each meta-lenslet bilayer.}
    \begin{tabular}{p{2cm} l l l l l l l l l l} \hline \hline
    \rule[-2ex]{0pt}{3.5ex} \multirow{3}{1.6cm}{\boldmath{$f_\alpha$} \bf{(mm)}} &\multicolumn{10}{c}{\bf{Numerical Aperture in Elongated Axis}}  \\ \cline{2-11}
    \rule[-1ex]{0pt}{3.5ex} & \multicolumn{10}{c}{Input Anamorphic Ratio}  \\ 
    & 1:1 & 1:2 & 1:3 & 1:4 & 1:5 & 1:6 & 1:7 & 1:8 & 1:9 & 1:10 \\ \hline
    \rule[-1ex]{0pt}{2ex} 2 & 0.10 & 0.20 & 0.30 & 0.40 & 0.50 & 0.60 & 0.70 & 0.80 & 0.90 & 1.0 \\
    \rule[-1ex]{0pt}{2ex} 3 & 0.07 & 0.13 & 0.20 & 0.27 & 0.33 & 0.40 & 0.47 & 0.53 & 0.60 & 0.67 \\ 
    \rule[-1ex]{0pt}{2ex} 4 & 0.05 & 0.10 & 0.15 & 0.20 & 0.25 & 0.30 & 0.35 & 0.40 & 0.45 & 0.50 \\
    \rule[-1ex]{0pt}{2ex} 5 & 0.04 & 0.08 & 0.12 & 0.16 & 0.20 & 0.24 & 0.28 & 0.32 & 0.36 & 0.40 \\
    \rule[-1ex]{0pt}{2ex} 6 & 0.03 & 0.07 & 0.1 & 0.13 & 0.17 & 0.20 & 0.23 & 0.27 & 0.30 & 0.33 \\
    \rule[-1ex]{0pt}{2ex} 7 & 0.03 & 0.06 & 0.09 & 0.11 & 0.14 & 0.17 & 0.20 & 0.23 & 0.26 & 0.29 \\ 
    \rule[-1ex]{0pt}{2ex} 8 & 0.03 & 0.05 & 0.08 & 0.10 & 0.13 & 0.15 & 0.18 & 0.20 & 0.23 & 0.25 \\ 
    \rule[-1ex]{0pt}{2ex} 9 & 0.02 & 0.04 & 0.07 & 0.09 & 0.11 & 0.13 & 0.16 & 0.18 & 0.20 & 0.22 \\ 
    \rule[-1ex]{0pt}{2ex} 10 & 0.02 & 0.04 & 0.06 & 0.08 & 0.10 & 0.12 & 0.14 & 0.16 & 0.18 & 0.20 \\ \hline \hline
    \end{tabular}
    \label{tab:NAf}
\end{table}

The meta-lenslets were modelled in OpticsStudio (Zemax). The layout consisted of two metasurfaces directly in contact with fused silica substrates, facing away from each other, as depicted in Figure~\ref{fig:LensletArray2}(b). There are four variable parameters: the phase profile of metasurface 1, the bilayer separation distance (air gap), the phase profile of metasurface 2, and the distance to the back focal plane. Since the 1:10 case is the most challenging to optimize for, the air gap and the back focal length were optimized for the 1:10 case. These distances were set as non-variable for all other anamorphic ratios between 1:1 - 1:9, so that each meta-lenslet in an array would satisfy parfocal operation.  

The $x$ and $y$ axes were chosen to be the elongated and the non-elongated axes, respectively. The anamorphic ratio was represented by the ratio of field angles in $x$ and $y$. For example, the 1:10 anamorphic ratio was represented by field angles of (0.3,0) and (0,0.03). To optimize for RMS spot size and anamorphic compression, the optimization routine included setting the centroid positions in the focal plane of the elongated and non-elongated fields (operands CENX and CENY) to be equidistant, but a higher weighting was given to requiring the RMS radius to be smaller than the Airy radius. The image size is quoted as the centroid position of the elongated field angle plus the radius of the RMS spot size.

The metasurfaces were modelled using `Binary 1' surfaces, which impart a phase delay according to ${x^2}$ and ${y^2}$ polynomial coefficients, as set out in Equation~\ref{eq:ana}. The normalized limits on the ${x^2}$ and ${y^2}$ coefficients are set to -250 to 250, reflecting the \qty{3}{\um} per $\mathrm{2\pi}$ phase shift limitation. Note that the polynomial is normalized by the radius of the meta-lenslet (\qty{200}{\um}) and is not expressed in terms of $\mathrm{2\pi}$.

To make a comparison with conventional optics, a spherical design was also modelled in OpticsStudio. The design parameters are held constant for both the spherical and meta-lenslet designs between each test to allow for a fair comparison. The spherical lenslet was optimized for the radius of the plano-convex lens and the back focal distance for the same fields as mentioned above, and the same 589~nm wavelength. The thickness of the silica substrate was set to 1.2~mm, a common thickness found in commercially available lenslet arrays. The diameter of the spherical lenslet was set to \qty{400}{\um}, the same as for the meta-lenslet. 

\subsection{\label{sec:OptResults}Optical System Design Results}

\begin{figure}
    \centering
    \includegraphics[width=\textwidth]{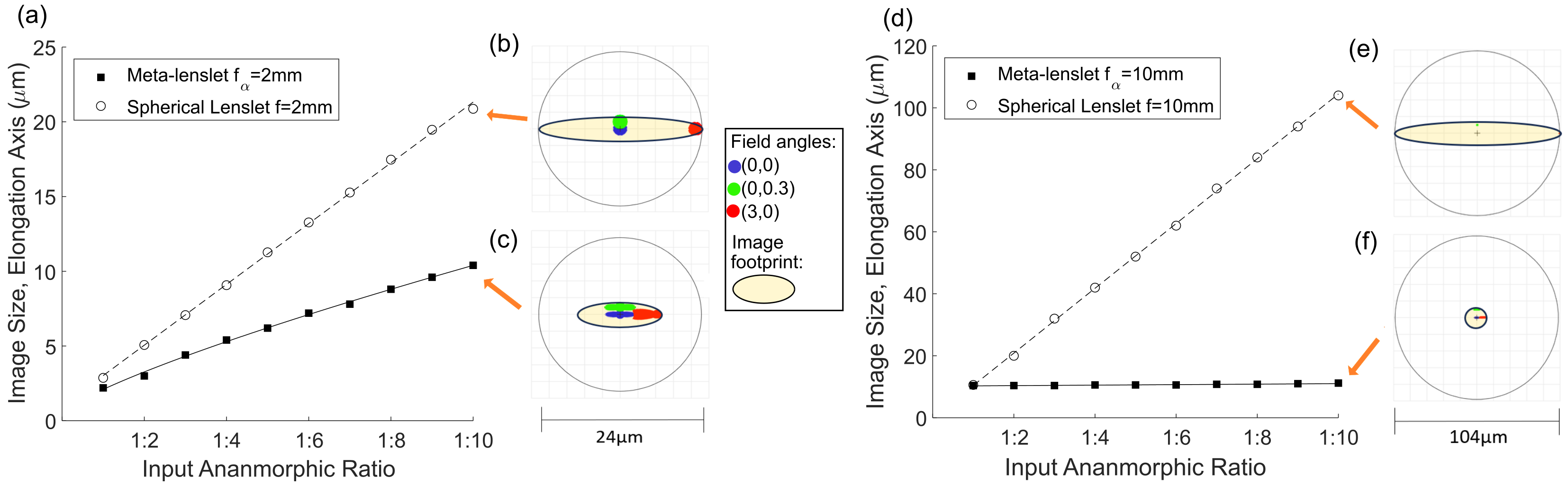}
    \caption{(a,d) Graph of image size versus input anamorphic ratio for a meta-lenslet and a spherical lenslet.
    The focal length is (a,b,c)~$\mathrm{f_\alpha}$ = \qty{2}{mm} and (d,e,f)~$\mathrm{f_\alpha}$ = \qty{10}{mm}.
    (b,c,e,f) Focal plane field angle positions and image footprint size at 1:10 anamorphism for (b,e)~spherical lenslet and (c,f)~meta-lenslet.
    The image size is reduced by (c)~58\% and (f)~98\% for a meta-lenslet relative to a spherical lenslet, as defined in Eq.~(\ref{eq:red}). 
    }
    \label{fig:TwoAndTen}
\end{figure}

For our design target case of $f_\alpha$ = \qty{2}{mm}, the image size is reduced by the meta-lenslet compared to the image size produced by the spherical lens across all input anamorphic ratios. Figure~\ref{fig:TwoAndTen}(a) shows a graph of the image size for each input anamorphic ratio for the meta-lenslet and spherical lenslet. Figures~\ref{fig:TwoAndTen}(b) and (c) show the image sizes for the spherical lenslet and the meta-lenslet for the 1:10 input anamorphic ratio, respectively. 

The spherical 1:1 anamorphic case defines the minimum image size, thereby the reduction in image size for the 1:10 anamorphic case is calculated as:
\begin{equation}
    \text{Reduction (\%)} = \frac{\text{meta-lenslet image size} - \text{minimum image size}}{\text{spherical image size} - \text{minimum image size}} * 100 .
    \label{eq:red}
\end{equation}
This scales the result so that a 100\% image size reduction corresponds to the elongated image being compressed down to the minimum image size, instead of zero. The image size is reduced by 94\% for the smallest anamorphic ratio of 1:2, and is between 58\% and 64\% for anamorphic ratios from 1:3 to 1:10. As predicted by the paraxial solutions to each case in Table~\ref{tab:NAf}, 1:3 to 1:10 were not able to be fully compressed due to the higher numerical apertures needed. There is inherently greater RMS spot spread in the elongation axis as the anamorphic ratio increases due to the increasing required numerical aperture. 

Since full image size reduction cannot be achieved for $f_\alpha$ = \qty{2}{mm}, counter-intuitively, it may be beneficial to enlarge the smallest images to be similar in size to most of the array to attain a more favourable plate scale; to test this assertion, the effect on the SNR variation and wavefront error would need to be quantified.  

Comparatively, for the longer focal length of $\mathrm{f_\alpha}$ = \qty{10}{mm}, full image size reduction is achieved across all input anamorphic ratios, as shown in Figure~\ref{fig:TwoAndTen}(d). Figures~\ref{fig:TwoAndTen}(e) and (f) show the image sizes for a spherical lenslet and the meta-lenslet for the 1:10 input anamorphic ratio, respectively. The greater RMS spot spread in the elongation axis as the anamorphic ratio increases, resulting in the meta-lenslet image size increasing from \qty{10.4}{\um} to \qty{11.2}{\um} for input anamorphic ratios 1:1 and 1:10, respectively.

In every case, the meta-lenslet was able to reduce the image size by compressing the image scale in the elongation axis. As predicted by the paraxial approximation, the effective focal length and anamorphic ratio combinations that required numerical apertures above 0.25 in the elongation axis were unable to be fully compressed. Table~\ref{tab:Resid} gives the percentage image size reduction for all cases. 

\begin{table}
    \centering
    \caption{Percentage image size reduction for each meta-lenslet bilayer, with varying input anamorphic ratio and effective focal length.}
    \begin{tabular}{p{2cm} l l l l l l l l l l} \hline \hline
    \rule[-2ex]{0pt}{3.5ex} \multirow{3}{1.6cm}{\boldmath{$f_\alpha$} \bf{(mm)}} &\multicolumn{10}{c}{\bf{Image Size Reduction (\%)}}  \\ \cline{2-11}
    \rule[-1ex]{0pt}{3.5ex} & \multicolumn{10}{c}{Input Anamorphic Ratio}  \\ 
    & 1:1 & 1:2 & 1:3 & 1:4 & 1:5 & 1:6 & 1:7 & 1:8 & 1:9 & 1:10 \\ \hline
    \rule[-1ex]{0pt}{2ex} 2 & - & 98 & 64 & 59 & 60 & 58 & 60 & 59 & 59 & 58 \\
    \rule[-1ex]{0pt}{2ex} 3 & - & 103 & 91 & 78 & 76 & 77 & 76 & 77 & 76 & 76 \\
    \rule[-1ex]{0pt}{2ex} 4 & - & 102 & 97 & 94 & 91 & 81 & 84 & 86 & 85 & 84 \\
    \rule[-1ex]{0pt}{2ex} 5 & - & 101 & 99 & 97 & 95 & 95 & 92 & 86 & 90 & 88 \\
    \rule[-1ex]{0pt}{2ex} 6 & - & 101 & 99 & 98 & 97 & 96 & 96 & 94 & 92 & 91 \\
    \rule[-1ex]{0pt}{2ex} 7 & - & 100 & 100 & 99 & 98 & 98 & 97 & 95 & 93 & 94 \\
    \rule[-1ex]{0pt}{2ex} 8 & - & 100 & 100 & 99 & 99 & 99 & 98 & 97 & 97 & 97 \\
    \rule[-1ex]{0pt}{2ex} 9 & - & 100 & 100 & 100 & 99 & 99 & 99 & 98 & 98 & 98 \\ 
    \rule[-1ex]{0pt}{2ex} 10 & - & 100 & 100 & 100 & 100 & 99 & 99 & 99 & 98 & 98 \\ \hline \hline
    \end{tabular}
    \label{tab:Resid}
\end{table}


\section{\label{sec:nanoResults}Meta-lenslet Nanostructure Modelling}

Our nanostructure design consists of $\mathrm{TiO_2}$ cylindrical pillars on a \qty{460}{\um} thick fused silica substrate. The unit cell is one cylindrical pillar on a defined area of substrate. To meet Nyquist sampling requirements, the unit cell size should be less than $\mathrm{{\lambda}/{(2\, NA)}}$~\cite{Khorasaninejad2017}. Given our design wavelength of \qty{589}{nm}, and taking a numerical aperture of 1, the unit cell should be less than \qty{295}{nm}. However, as shown in Section~\ref{sec:OpticalDesigns}, we are expecting to only utilize maximum numerical apertures of $\mathrm{NA}\sim$~0.3. Therefore, we obtain a corresponding unit cell size of less than \qty{982}{nm}. In our design, we chose a unit cell size of \qty{350}{nm}, corresponding to a numerical aperture of 0.8, thereby giving adequate sampling well above our expected maximum numerical aperture while also having acceptable transmission~\cite{Ozdemir2018}.

To simulate the electric and magnetic field components inside the nanostructures, we use a rigorous coupled wave analysis implemented in the RETICOLO software package~\cite{Hugonin:2101.00901:ARXIV}. We calculate the phase shift and transmission response for a periodic arrangement of cylinders of heights varying in the range of \qtyrange{300}{650}{nm}, with \qty{50}{nm} increments. The diameter of the cylinders was also varied between \qtyrange{0}{350}{nm} with approximately \qty{4}{nm} increments.

\begin{figure}[t]
    \centering
    \includegraphics[width=\textwidth]{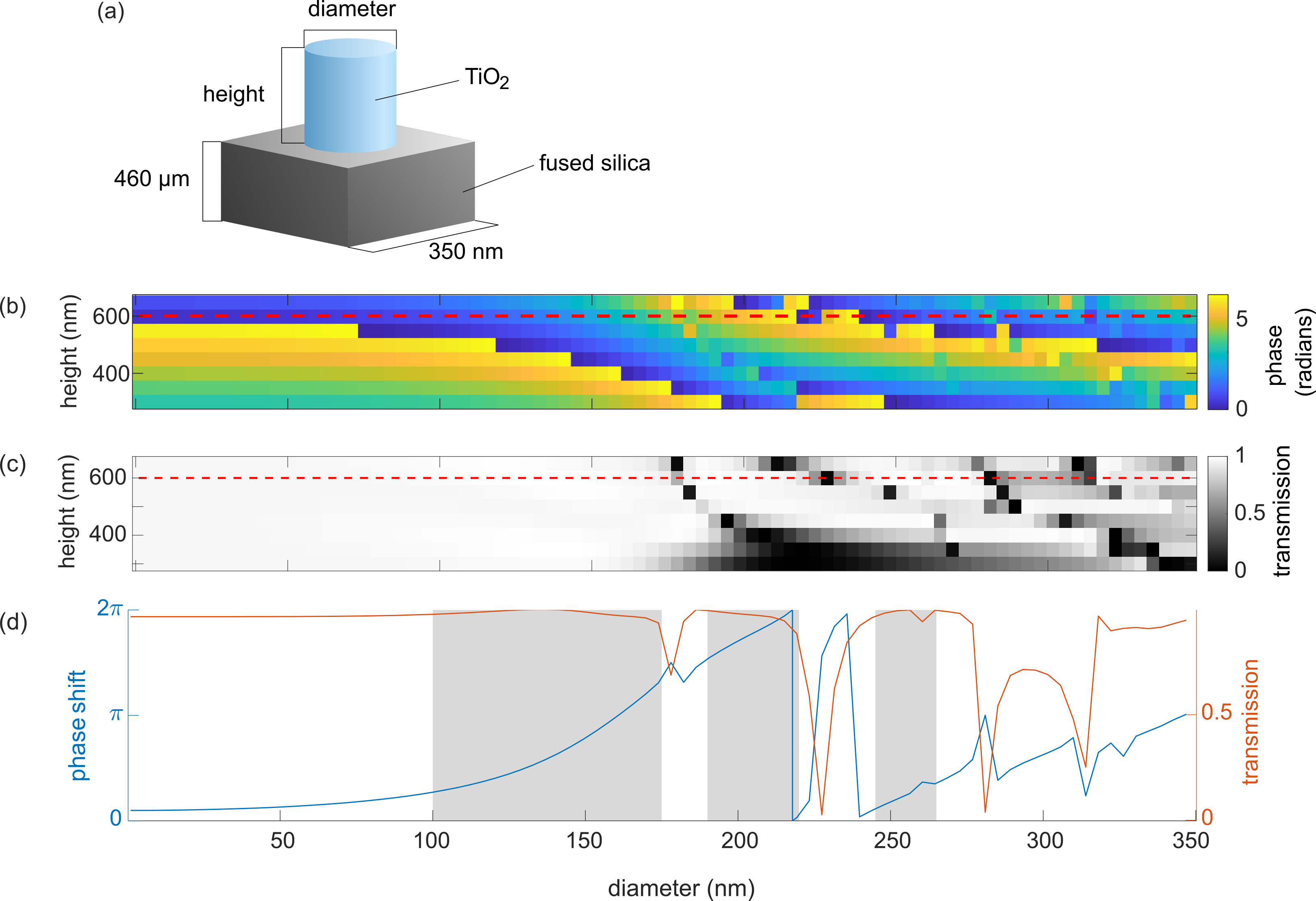}
    \caption{
    (a) Diagram of the metasurface unit cell, with a \qty{350}{nm} square fused silica substrate, and a $\mathrm{TiO_2}$ cylindrical pillar. 
    (b)~Transmission (normalised intensity) and (c)~phase shift calculated for varying diameters and cylinder heights. 
    (d)~Phase shift and transmission (normalised intensity) versus diameter of the \qty{600}{nm} high pillars. The grey shaded areas indicate regions of pillar diameter that have acceptable transmission and are above the minimum diameter required for fabrication. There is a full $\mathrm{2 \pi}$ shift accessible in these regions.}
    \label{fig:PhaseTrans}
\end{figure}

As shown in Figure~\ref{fig:PhaseTrans}(a), the unit cell was modelled as a $\mathrm{TiO_2}$ cylindrical pillar with a \qty{350}{nm} $\times$ \qty{350}{nm} $\times$ \qty{460}{\um} fused silica substrate. The effect of the cylindrical pillar height on the transmission (normalised intensity) and phase delay results are shown in Figures~\ref{fig:PhaseTrans}(b) and (c), respectively. Results of cylinder diameters below \qty{50}{nm} are shown for completeness, but they are not included in the full $\mathrm{2\pi}$ phase coverage considerations, as pillars of these sizes cannot be adequately fabricated. Although a full $\mathrm{2\pi}$ phase shift is accessible for all of the cylinder heights tested, there is significant transmission loss for the short cylinders with heights $<400$~nm. 

As a result of these parameter maps, we select a design based on a cylinder with a height of \qty{600}{nm}. This choice of height gives a full $\mathrm{2\pi}$ phase shift in regions of high transmission for cylinders with diameters between $\sim$\qtyrange{100}{260}{nm}, as shown by the grey regions in Figure~\ref{fig:PhaseTrans}(d). A metasurface within these design constraints is expected to be comfortably within acceptable levels of fabrication accuracy and repeatability using current nanofabrication technology.  

We then construct the meta-lenslets by searching for the meta-atom in our library that produces the smallest phase offset from the target phase at each spatial location. 
Since each of our meta-lenslets have mirror symmetry in both the $x$ and $y$ axes, only one quadrant of the device is constructed, which is then translated and mirrored to build the full device. This approach saves computation time by a factor of four.  

\begin{figure}
    \centering
    \includegraphics[width=\textwidth]{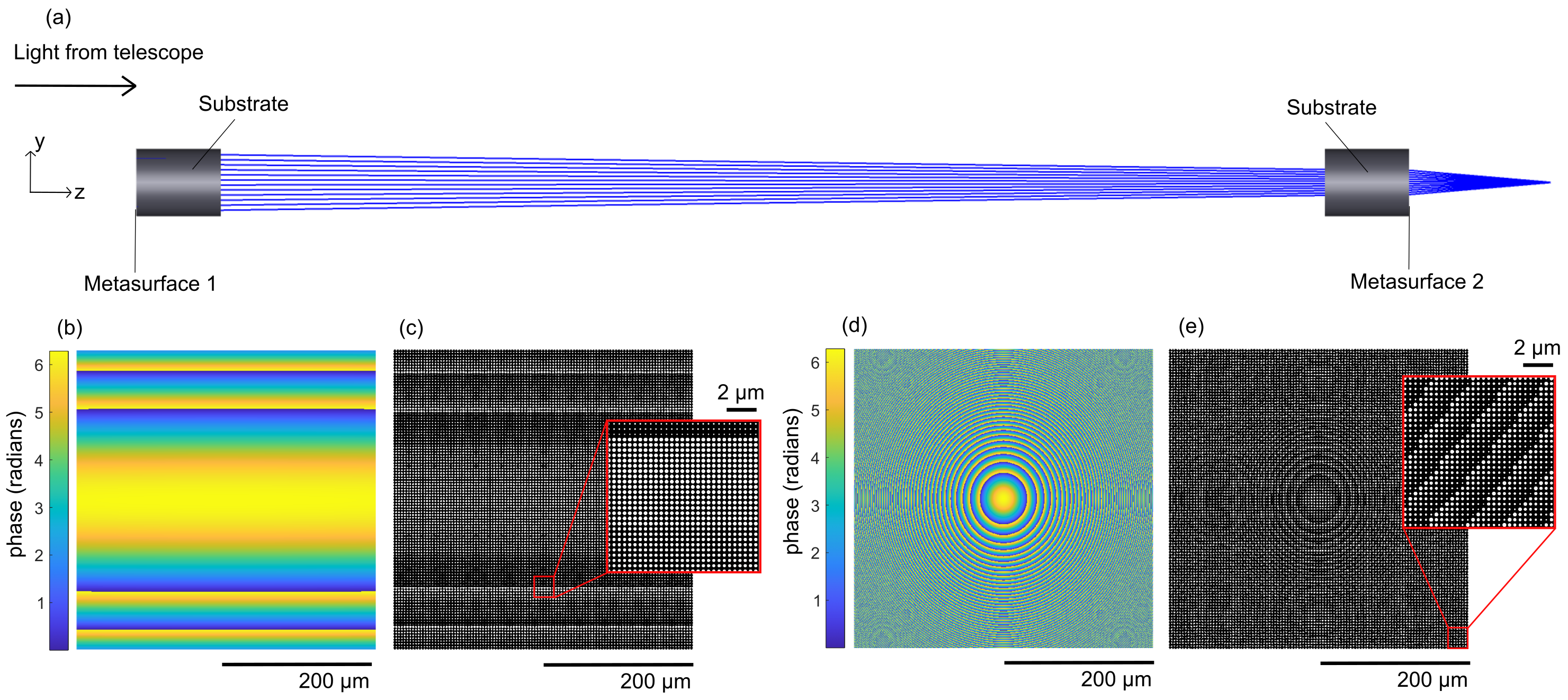}
    \caption{
    (a) Optical layout of a meta-lenslet bilayer with $\mathrm{f_\alpha}$ = \qty{2}{mm}, and a 1:10 anamorphic ratio, as modelled in OpticsStudio. 
    (b,d)~Phase profile generated from the OpticsStudio model for (b)~metasurface 1, and (d)~metasurface 2. 
    (c,e) $\mathrm{TiO_2}$ cylinders of varying diameters impart the required phase profile for (c)~metasurface~1 and (e)~metasurface~2. The zoomed-in areas show the varying diameters of the pillars in an area of large phase shift. }
    \label{fig:devices}
\end{figure}

The meta-lenslet device design for the $\mathrm{f_\alpha}$ = \qty{2}{mm}, 1:10 anamorphic ratio case is shown in Figure~\ref{fig:devices}. The first metasurface imparts a mostly cylindrical phase profile [Figs.~\ref{fig:devices}(b,c)], while the second metasurface imparts a slightly toroidal profile with stronger focusing in the $x$ axis than the $y$ axis [Figs.~\ref{fig:devices}(d,e)]. The edges of the second metasurface in this design are the steepest changes in phases of all the devices simulated. It can be seen that even in this most extreme example, there are approximately eight pillars covering a full $\mathrm{2\pi}$ change, supporting the feasibility of the design. 

\section{\label{sec:Conclusion}Conclusion and Future Work}

Prior studies have noted the importance of decreasing the detector size for LGS wavefront sensing for ELTs. This study explored the feasibility of using a meta-lenslet array to anamorphically compress the variable LGS image size on a wavefront sensor to allow a decrease in detector size. 

We have shown a conceptual design for a meta-lenslet array with an effective focal length of $\mathrm{f_\alpha}$ = \qty{2}{mm} to match the current design parameters for GMT's LTAO. The meta-lenslets do not compress the higher anamorphic ratios fully at this focal length, and consequently have some residual laser guide star image elongation, but still offer significant improvement over the static plate-scale problem as a reduction of \qty{50}{\percent} in image size for the largest anamorphic compressions was achieved. However, we have also shown that effective focal lengths $\mathrm{f_\alpha \gtrsim}$~\qty{8}{mm} would be capable of complete anamorphic compression across a range of 1:1-1:10 anamorphic ratios. Noting that the current design for GMT's LTAO is to compromise the pixel scale to adequately sample smaller images and avoid excessive truncation of larger LGS images, relaxing the focal length requirements to achieve full anamorphic compression across the whole SHWFS could be considered. 

Future work will focus on fabricating a prototype to verify the physical parameters and performance of meta-lenslets with a range of anamorphic compression ratios. A design validation prototype can be made with silicon, as the fabrication process is mature and reliable at the expense of throughput efficiency at the target visible wavelength. Further designs can employ $\mathrm{TiO}_2$ to gain a higher throughput efficiency.


\appendix{}    

\subsection*{Disclosures}

The authors have no conflicts of interest to declare.

\subsection* {Code, Data, and Materials Availability} 
Code and data can be made available upon reasonable request to the corresponding author.

\subsection* {Acknowledgments}
We acknowledge the funding support of the Australian Research Council (NI210100072, CE200100010). This research was initially presented as a poster, "Meta-lenslet array for laser guide star anamorphic compression", by co-lead author S.E. Dean, Paper 12990-108, at SPIE Photonics Europe, Strasberg, France, 7-11 April 2024.  


\bibliography{spiebib}   
\bibliographystyle{spiejour}   

\end{document}